\documentclass[a4paper]{article}

\usepackage[left=1in,right=1in,top=1in,bottom=1in]{geometry}
\usepackage{supertabular}
\usepackage{amsthm}
\usepackage{amsfonts}
\usepackage{subfig}
\usepackage{amsmath}
\usepackage{amssymb}
\usepackage{graphicx}
\graphicspath{{Fig/eps/}}

\def\longrightharpoonup{\relbar\joinrel\rightharpoonup}
\def\longleftharpoondown{\leftharpoondown\joinrel\relbar}

\def\longrightleftharpoons{
  \mathop{
    \vcenter{
      \hbox{
      \ooalign{
        \raise1pt\hbox{$\longrightharpoonup\joinrel$}\crcr
	  \lower1pt\hbox{$\longleftharpoondown\joinrel$}
	  }
      }
    }
  }
}

\newcommand{\rates}[2]{\displaystyle
\mathrel{\longrightleftharpoons^{#1\mathstrut}_{#2}}}

\begin{document}
\raggedbottom

\title{Variance reduction for steady-state simulation and sensitivity analysis of stochastic chemical systems}

\author{Andreas Milias-Argeitis \thanks{Department of Biosystems Science and Engineering, ETH Zurich, 4058 Basel, Switzerland; andreas.milias@bsse.ethz.ch} \and John Lygeros \thanks{Automatic Control Lab, ETH Zurich, Physikstrasse 3, 8092, Zurich, Switzerland; lygeros@control.ee.ethz.ch} \and Mustafa Khammash \thanks{Department of Biosystems Science and Engineering, ETH Zurich, 4058 Basel, Switzerland; mustafa.khammash@bsse.ethz.ch}}

\maketitle

\begin{abstract}
We address the problem of estimating steady-state quantities associated to systems of stochastic chemical kinetics. In most cases of interest these systems are analytically intractable, and one has to resort to computational methods to estimate stationary values of cost functions. In this work we consider a previously introduced variance reduction method and present an algorithm for its application in the context of stochastic chemical kinetics. Using two numerical examples, we test the efficiency of the method for the calculation of steady-state parametric sensitivities and evaluate its performance in comparison to other estimation methods.
\end{abstract}

\section{Introduction}
Knowledge of steady-state quantities related to an ergodic stochastic chemical system can provide useful insights into its properties. Moreover, steady-state values of cost functions are often easier to estimate with good accuracy compared to stationary distributions. When the system propensities are affine in the state, mean values of polynomial functions of the system state can be computed analytically, as the system of moments is closed. However, when non-polynomial functions are considered, or the system propensities are not affine, analytic calculations are no longer possible and the only solution left is simulation.

While moment closure methods \cite{singh11} can be used to provide good approximations to system moments over a finite time interval, they commonly tend to diverge from the true solution over time, thus resulting in biased steady-state values. The solution presented in Ref. \cite{ruess11} works only when polynomial functions of the state are considered, and generalization to arbitrary functions is still very difficult. The Finite State Projection algorithm \cite{munsky06} can be alternatively employed to provide moment estimates with guaranteed accuracy bounds, however the number of states required to attain a certain accuracy makes the method applicable to small problems.

On the other hand, stochastic simulation \cite{gillespie07} can always provide estimates for the stationary mean of any function of the state, however these estimates are inevitably noisy. Brute-force noise reduction can only be achieved at an increased computational cost, either by simulating longer trajectories or by running many trajectories in parallel.

Another possibility for reducing the noise in the estimated quantities is the application of a variance reduction technique \cite{asmussen}, provided the added computational cost of the reduced variance estimator is significantly smaller than the gain in computer time. In this work we present the application of such a variance reduction technique \cite{henderson1,henderson2} to systems of stochastic chemical kinetics. The idea is based on so-called \emph{shadow functions} and originated in the queueing systems simulation literature, a field where the range of analytically tractable systems in that field is much larger. We demonstrate how the same idea can be applied to steady-state simulation of stochastic chemical systems. We further test the capabilities of the reduced-variance estimators by performing parametric sensitivity calculations for two systems governed by nonlinear propensity functions.

The paper is organized as follows: in Sections II and III we define the steady-state estimation problem and define the na\"{i}ve and shadow function estimators. In Sections IV and V we present one possible implementation of the variance reduction technique to stochastic chemical kinetics and its applicability to steady-state sensitivity analysis. The numerical examples in Section VI serve to demonstrate the effectiveness of the shadow function method in practice and assess its computational cost in comparison to na\"{i}ve estimation. The conclusions of our study and some future research directions  are finally summarized in Section VII.

\section{Problem statement}
\subsection{Setup}
Assume an irreducible positive recurrent Markov chain $X=\{X(t):t\geq 0\}$ on a countable space $\mathcal{S}$. In the case of stochastic chemical kinetics, $\mathcal{S}\subseteq\mathbb{Z}^N_{\geq 0}$, where $N$ is the number of chemical species in the system. The chain moves according to a finite set of available transitions $\{\zeta_r\in \mathcal{S}\}_{r=1}^R$, with a corresponding set of propensity functions $\{\lambda_r:\mathcal{S}\to\mathbb{R}\}_{r=1}^R$. The infinitesimal generator of $X$ is the operator $Q$ satisfying
\begin{equation}\label{generator}
(Qf)(x) = \sum_{r=1}^R\lambda_r(x)(f(x+\zeta_r)-f(x))
\end{equation}
for all $f:\mathcal{S}\to\mathbb{R}$ such that $|Qf(x)|<\infty~\forall x\in \mathcal{S}$. The discreteness of $\mathcal{S}$ allows us to enumerate its elements and think of $Q$ as an infinite matrix $Q=(Q_{ij}),~i,j\in\mathbb{N}$. Similarly, any function $f$ on $\mathcal{S}$ can be thought of as an infinite column vector, and distributions on $\mathcal{S}$ can be defined as infinite row vectors.

\subsection{Steady-state estimators}
Let $f:\mathcal{S}\to\mathbb{R}$ be a $\pi$-integrable cost function associated with $X$. The ergodic theorem for Markov chains \cite{norris98} ascertains that for any initial condition \[\lim_{t\to\infty}\frac{1}{t}\int_0^t f(X(s))\,ds=\pi f:=\alpha~~\mbox{almost surely,}\] where $\pi$ is the unique invariant distribution of the system and $\alpha$ the steady-state mean value of $f$. Since the analytic calculation of $\alpha$ is possible only in very special cases, its estimation from simulation is usually the only possibility. The most straightforward estimator of $\alpha$ is \begin{equation}\alpha_1(t)=\frac{1}{t}\int_0^t f(X(s))\,ds,\end{equation} which is also strongly consistent \cite{asmussen}.

Under some further general conditions on $X$, and $f$, we also know that
\begin{equation}\label{CLT}
\sqrt{t}(\alpha_1(t)-\alpha)\Rightarrow \sigma_1\mathcal{N}(0,1),
\end{equation}
as $t\to\infty$, where $\Rightarrow$ denotes weak convergence, $\mathcal{N}(0,1)$ is  the standard normal distribution and $\sigma^2_1$ is called the \emph{time average variance constant (TAVC) for $\alpha_1(t)$} \cite{asmussen}.

The TAVC can be expressed in terms of the integrated autocovariance function of the process $(f_c(X(t)):t\geq 0)$, where $f_c(x):=f(x)-\alpha$, according to the formula \cite{asmussen}:
\begin{equation}\sigma_1^2=2\int_0^\infty\mathbb{E}_{\pi}[f_c(X(0))f_c(X(s))]\,ds.\end{equation}
An alternative expression for $\sigma^2$ can be derived from the functional Central Limit Theorem for continuous-time Markov chains \cite{bhattacharya}:
\begin{equation}\label{cltvar}
\sigma_1^2=-2\langle Qg,g\rangle=-2\int g(x)\cdot Qg(x)\,d\pi(x),
\end{equation}
where $g$ is a solution to the so-called \emph{Poisson's equation}\cite{asmussen2} (Note that solutions to the Poisson equation are unique up to an additive constant, i.e. if $g$ is a solution, then $g'=g+c,~c\in\mathbb{R}$ is also a solution \cite{asmussen2}):
\begin{equation}
\label{poissonexact}
Qg=-f_c.
\end{equation}

A more general class of estimators for $\alpha$ has the form
\begin{equation}\label{a2}\alpha_2(t)=\frac{1}{t}\int_0^t (f+h)(X(s))\,ds,\end{equation}
where $h:\mathcal{S}\to\mathbb{R}$ is chosen such that $t^{-1}\int_0^t h(X(s))\,ds\to 0$ almost surely for all $x\in \mathcal{S}$ \cite{henderson1}. The function $h$ offers an extra degree of freedom in the design of the estimator, which can be exploited to achieve variance reduction. In other words, $h$ can be chosen such that the TAVC of $\alpha_2$, denoted by $\sigma^2_2$, is smaller than $\sigma_1^2$. The obvious choice $h^{opt}=\alpha-f$ is of course intractable, however it suggests that a function $h$ with a zero steady-state mean that is approximately equal to $\alpha-f$ could also achieve variance reduction. Such functions would result in a process $h(X(\cdot))$ that behaves almost antithetically from $f(X(\cdot))$, thus making the variance of $(f+h)(X(\cdot))$ smaller than that of $f(X(\cdot))$ alone. In the steady-state simulation literature, a function $h:\mathcal{S}\to\mathbb{R}$ that satisfies $\pi h=0$ is called a \emph{shadow function} \cite{henderson2}.

The problem then becomes the selection of an appropriate shadow function $h$, so that $\sigma_2^2=c\cdot\sigma_1^2$, with $c<1$. From \eqref{CLT} we see that a reduction of variance by a factor $c$ implies that the variance of $\alpha_2(T)$ is equal to the variance of $\alpha_1(T/c)$. Assuming that the computational cost of both estimators is dominated by the cost of simulating the process $X$, $c^{-1}$ can be used as an indicator of the efficiency of $\alpha_2$ relative to $\alpha_1$.

The basic idea of the shadow function method of Ref. \cite{henderson2}, outlined in the next section, is to obtain such an $h$ by using analytical information from a second Markov chain that approximates the original one and is mathematically tractable. A second alternative solution of more general applicability will be described after presenting the method in more detail.

\section{The shadow function method: main idea}
The basic idea to the shadow function method is to consider candidate functions of the form \begin{equation}h=Qg,\end{equation} where $Q$ is the generator matrix of the Markov chain and $g$ is any $\pi$-integrable function (so that the ergodic theorem holds for it as well). In this case, and under the assumption that $\pi(Qg)=(\pi Q)g$ (that holds under some not-too-stringent conditions on $g$ \cite{henderson2}), $Qg$ becomes a shadow function. We are then naturally led to consider the solution of the Poisson equation \eqref{poissonexact},
which could provide us with the appropriate function $g$. Solving \eqref{poissonexact} is of course not possible, since the state space is countable and $\alpha$ is unknown. However, we can look for so-called \emph{surrogate functions} that approximate this solution to build a better estimator.

Following the analysis from \cite{henderson2}, we consider another Markov chain $\tilde{X}$ evolving on a countable space $\tilde{\mathcal{S}}$, with stationary distribution $\tilde{\pi}$ and generator $\tilde{Q}$. We also assume a map $r:\mathcal{S}\to\tilde{\mathcal{S}}$ (not necessarily one-to-one) and a function $\tilde{f}$ that is somehow closely related to the original cost function $f$. If $\tilde{f}$ is $\tilde{\pi}$-integrable, we further assume that we can compute the solution to the Poisson equation
\begin{equation}
\label{poissonappr}
\tilde{Q}\tilde{g}=\tilde{\pi}\tilde{f}-\tilde{f},
\end{equation}
through which we arrive at a surrogate function \begin{equation}\label{newg}g(x)=\tilde{g}(r(x))~\forall x\in \mathcal{S}.\end{equation}

Summing up, the approach outlined above is based on the fact that if $\tilde{X}$ is: 1) a relatively good approximation of $X$ and 2) tractable analytically, then we can derive a surrogate function $g$ and an estimator $\alpha_2(t)$ which is better than the original estimator in terms of TAVC (assuming that the extra calculation time needed for $\alpha_2$ is not significant).

\section{Practical implementation of the shadow function method in chemical kinetics}
The shadow function method was originally developed for steady-state simulation of queueing systems, for which a wide range of known and tractable approximations exists. The solution of the approximating Poisson equation can thus be calculated explicitly in many cases, and the application of the method is straightforward. This is not the case for stochastic chemical kinetic systems, where explicit solutions are very hard or impossible to calculate. One thus has to resort to different types of approximation schemes, outlined below.

\subsection{State-space truncation}
The Markov chains we are interested in satisfy the following properties:
\begin{enumerate}
\item They have a finite number of bounded increments over each finite time interval
\item Each state leads to a finite number of states (i.e. for every $i$, $Q(i,j)\neq 0$ for finitely many $j$'s)
\end{enumerate}
For such chains, an obvious idea for obtaining an approximating process is to consider a chain evolving on a finite truncation of $\mathcal{S}$ (i.e. consider $\tilde{\mathcal{S}}$ to be a finite subset of $\mathcal{S}$). Actually, under quite weak assumptions and careful definition of $\tilde{Q}$, one can show that the invariant distribution of $\tilde{X}$ on $\tilde{\mathcal{S}}$ approaches that of $X$ as the truncation size grows \cite{tweedie98}. This of course implies that $\tilde{\pi}\tilde{f}$ also approaches $\pi f$. In this case, the function $r$ between the two state spaces can be intuitively defined to map every $s\in \mathcal{S}\bigcap\tilde{\mathcal{S}}$ to itself, and every $s\in \mathcal{S}\setminus\tilde{\mathcal{S}}$ to some $\tilde{s}\in\tilde{\mathcal{S}}$ (which may vary with $s$). In this way, $\tilde{f}=f|_{\tilde{\mathcal{S}}}$.

In order to arrive at a good approximation with this approach, one first has to study a few simulations of $X$, to determine a finite set that contains a good amount of its invariant mass and then perform the necessary calculation of the solution to the Poisson equation on $\tilde{\mathcal{S}}$. The size of this set is determined in practice as a trade-off between tractability and approximation accuracy. However, the applicability of this approach is in general very limited due to the fact that the required truncations grow exponentially with the system dimension. Another problem is that the approximation $\tilde{g}$ of $g$ (the solution to the original intractable Poisson equation) will be very poor for states $s\in \mathcal{S}\setminus\tilde{\mathcal{S}}$, because of the form of $r$, which projects are states outside $\tilde{\mathcal{S}}$ back into the set. This implies that significant variance reduction will be hard to achieve (and in some cases variance may even increase), if the chain sample paths exit $\tilde{\mathcal{S}}$ too frequently during simulation.

\subsection{Approximating solutions of the Poisson equation}
Instead of searching for an approximating Markov process, one may try to approximate the solution of \eqref{poissonexact} directly, to arrive at a suitable shadow function $h$. This approach is also followed in Ref. \cite{meyn07}, where the discrete-time steady-state simulation problem is considered. Given a set of functions $\{\psi_i:\mathcal{S}\to\mathbb{R},~i=1,\dots,n\}$ \footnote{Candidate functions $\psi_i$ must satisfy a boundedness condition derived from a Foster-Lyapunov inequality. For more details, see Ref. \cite{glynn96} or Ch.8 of Ref. \cite{meyn07}. In the sequel we will assume that all the functions considered satisfy this property.}, one can define
\begin{equation}\label{ghat}
\hat{g}=\sum_{i=1}^n\theta_i\psi_i=\psi\cdot\theta,
\end{equation}
where $\psi=\begin{bmatrix}\psi_1\dots\psi_n\end{bmatrix}$ and $\theta\in\mathbb{R}^{n\times 1}$ is a vector of weights.

In principle one could then try to calculate the value of $\theta$ that minimizes the TAVC of $\alpha_2$. Using \eqref{cltvar} and \eqref{ghat}, this variance constant turns out to be (see Appendix \ref{app_A})
\begin{equation}\label{theta_opt}
\sigma_2^2=\sigma_1^2-2\left[\langle f_c,\psi\theta\rangle -\langle Q(\psi\theta),g\rangle +\langle Q(\psi\theta),\psi\theta\rangle\right],
\end{equation}
where $g$ solves \eqref{poissonexact}. Thus, minimizing the TAVC of $\eqref{a2}$ requires knowledge of $g$, which is unavailable.

We thus have to resort to heuristic methods for obtaining a suboptimal estimate of $\theta$, for example by determining the value of $\theta$ that minimizes
\[L(\theta):=\int(Q(\psi\theta)+f_c)^2\,d\mu(x)\]
for some suitable measure $\mu$. This is a linear least squares regression problem, which can be solved approximately by generating a set of training data $(f_c(x_1),Q\psi(x_1)),\dots,(f_c(x_m),Q\psi(x_m))$, $x_1,\dots,x_m$, with weights $\mu(x_1),\dots,\mu(x_m)$.

If we define the finite-sample version of $\psi$ by
\[\Psi:=\begin{bmatrix}\psi_1(x_1)&\psi_2(x_1)&\dots&\psi_n(x_1)\\\psi_1(x_2)&\psi_2(x_2)&\dots&\psi_n(x_2)\\ \dots&\dots&\dots&\dots\\ \psi_1(x_m)&\psi_2(x_m)&\dots& \psi_n(x_m)\end{bmatrix}\]
and similarly set
\[F_c:=\begin{bmatrix}f_c(x_1)&f_c(x_2)&\dots&f_c(x_m)\end{bmatrix}^T,\]
we can then calculate the matrix $\Psi_Q\in\mathbb{R}^{m\times n}$ corresponding to $Q\psi$ by using the explicitly known form of the Markov chain generator \eqref{generator} and finally obtain
\begin{equation}
\theta^*=(\Psi_Q^TM\Psi_Q)^{-1}\Psi_Q^TMF_c,
\end{equation}
where $M=diag(\mu(x_1),\dots,\mu(x_m))$, as the (weighted) least squares minimizer of $L(\theta)$.

\subsection{Variance reduction algorithm using a shadow function}
Putting together all the elements presented above, we summarize below the basic steps of the variance reduction algorithm implemented in this work:
\begin{itemize}
  \item[(1)]Simulate a long path of the process $X$ using any preferred version of the stochastic simulation algorithm \cite{gillespie07}.
  \item[(2)]Obtain a rough estimate of $\alpha$ from the simulated trajectory using $\alpha_1$.
  \item[(3)]Pick a set of functions $\psi_i,~i=1,\dots,n$ and approximate the solution $g$ to the Poisson equation \eqref{poissonexact} by $\hat{g}=\psi\cdot\theta^*$ using the approach outlined above.
  \item[(4)]Evaluate $h=Qg$ along the simulated sample path.
  \item[(5)]Refine the estimate of $\alpha$ using $\alpha_2$.
  \item[(6)]Verify that variance reduction has been achieved.
\end{itemize}

The last step is necessary to ensure that the variance has not actually increased due to the use of a suboptimal weight vector $\theta$, and it can be carried out quite straightforwardly using the method of batch means \cite{asmussen} and the simulated trajectory from Step 1. In all cases we have tested, Steps 2-6 do not contribute more than a few seconds to the computational cost of this algorithm, which implies that the main computational bottleneck still lies at Step 1.

\subsubsection{Implementation issues}\label{implementation}
The estimate of $\theta^*$ obtained by weighted least squares is clearly suboptimal, however it may still yield a reduced-variance estimator. The choice of the weighting measure $\mu$ in the optimization problem above is completely free, and one could in principle try to optimize over both $\mu$ and $\theta$ for a given problem. In practice however, such an approach would increase computational cost of the reduced-variance estimator and possibly eliminate the benefit of variance reduction. To maintain estimator efficiency, one should thus consider a single (or a few) ``generic'' choices for $\mu$, and preferably re-use the points generated at Step 1.

A reasonable choice of weighting measure would be $\pi$ itself. The training set for regression would then consist of all distinct points visited by the process over the course of simulation in Step 1 (possibly after discarding the burn-in period), weighted according to the empirical distribution of the process. A more coarse approximation of $\pi$ would be to use the same sample with all weights being equal. Yet another possibility consists of sampling from a uniform grid that is centered on the area containing the bulk of the invariant mass of the chain. This area can also be crudely determined from the sample of Step 1. All these approaches can achieve variance reduction, however the optimal choice remains problem-dependent. Given that the calculation of least squares estimates can be carried out very efficiently using linear algebraic techniques, it is highly advisable to test several alternatives for the problem at hand.

In Ch.11 of Ref. \cite{meyn07}, the problem of selecting an optimal $\theta$ is overcome by introducing a least-squares temporal difference learning (LSTD) algorithm for the approximation of the value of $\theta$ that minimizes the variance of $\alpha_2$ in the context of discrete-time chains. The same algorithm could in principle be applied to continuous-time chains using the embedded discrete-time Markov chain and carrying out the necessary modifications to the original algorithm, based on the results of Ref. \cite{hordijk}. While this solution is theoretically justified, it requires setting up and running an LSTD estimator in parallel with the simulated chain that will asymptotically converge to the optimal value of $\theta$. Depending on the convergence properties of this estimator, the overall efficiency of the variance reduction scheme may be smaller than the efficiency achieved by using a sub-optimal value for $\theta$, especially when several approximating functions $\psi_i$ are considered.

Another degree of freedom in the design of shadow function estimators is the choice of the approximating set $\{\psi_i,i=1,\dots,n\}$. Here, the probabilistic interpretation of Poisson's equation may assist the selection of approximating functions by providing some useful intuition: Assuming $f$ is $\pi$-integrable and $X$ ergodic, it holds that \cite{asmussen2,makowski02}
\[g(x)=\mathbb{E}_x\left[\int_0^{\tau(x_0)}f_c(X_s)\,ds\right],\]
where $\tau(x_0)$ is the hitting time of some state $x_0$ (changing $x_0$ simply shifts $g(x)$ by a constant) and $\mathbb{E}_x$ denotes expectation given $X(0)=x$. From this equation one may infer some general properties of $g$ (e.g. monotonicity, oscillatory behavior etc.) based on the form of the propensity functions. The same formula can be used to provide some crude simulation-based estimates of $g(x)$, which can be also helpful for the selection of $\{\psi_i\}$. Finally, a Lyapunov-type analysis can be employed to infer the asymptotic behavior of $g$ \cite{glynn96}.

\section{Steady-state parameter sensitivity}
Chemical reaction systems typically depend on several kinetic parameters, and the calculation of the output sensitivity with respect to these parameters is an essential step in the analysis of a given model. While there are several powerful parameter sensitivity methods available today \cite{sheppard12, anderson12}, they are mostly appropriate for transient sensitivity analysis, as the variance of their estimates tends to grow with the simulation length. Indeed, it can be shown that the variance of sensitivity methods based on the so-called likelihood ratio \cite{glynn90} or the Girsanov transformation \cite{plyasunov07} grows linearly with time. On the other hand, the variance of estimators based on finite parametric perturbations can be shown to remain bounded under mild conditions on the propensity functions, provided the underlying process is ergodic. However, the stationary variance can be still quite large, which makes necessary the use of a variance reduction method, such as the one presented here. Besides providing reduced-variance estimates of various steady-state functions of the chain, the shadow function estimator can be also employed for sensitivity analysis using a finite difference scheme \cite{asmussen} and the Common Random Numbers (CRN) estimator \cite{rathinam10}.

More analytically, assuming that the propensity functions of $X$ are of the form $\lambda(x,p)$, where $p$ is a parameter of interest, the finite difference method aims to characterize the sensitivity of the steady-state value of a given function $f$ to a small finite perturbation of $\delta$ of $p$ around a nominal value $p_0$. If $\delta$ is small enough, we expect that $(\alpha(p_0+\delta)-\alpha(p_0))/\delta$ will be approximately equal to $\partial\alpha/\partial p$.

Finite difference-based sensitivity analysis using shadow functions can be simply carried out by generating process trajectories for the nominal and perturbed parameter values, and estimating $\partial\alpha/\partial p$ by $(\alpha_2(p_0+\delta)-\alpha_2(p_0))/\delta$. As shown in Ref. \cite{rathinam10}, use of the same random number stream for the generation of both the nominal and perturbed trajectories can result in great variance decrease compared to using independent streams.

\section{Numerical Examples}
To demonstrate the efficiency of shadow function estimators, we next present two applications of the method to steady-state sensitivity estimation. We compare our finite difference scheme that uses common random numbers and the shadow function estimator to the method of Coupled Finite Differences (CFD) \cite{anderson12}, which frequently outperforms finite-difference estimators based on common random numbers and the Random Time Change representation \cite{anderson12,rathinam10}.

All numerical examples were generated using custom-written Matlab scripts running on a 3.4 Ghz quad-core PC with 8 GB of RAM.
\subsection{Stochastic focusing}
As a first example, we consider the stochastic focusing model of \cite{paulsson00}, where an input signaling molecule $S$ inhibits the production of another molecule $R$. Stochastic focusing arises due to the presence of stochastic fluctuations in $S$, that make the mean value of $R$ more sensitive to changes $S$ than predicted by the deterministic model of the system. The same system is treated in Ref. \cite{warren12} using a more sophisticated method based on trajectory reweighting.

\begin{figure}[h!bt]
\centering
\includegraphics[width=\columnwidth]{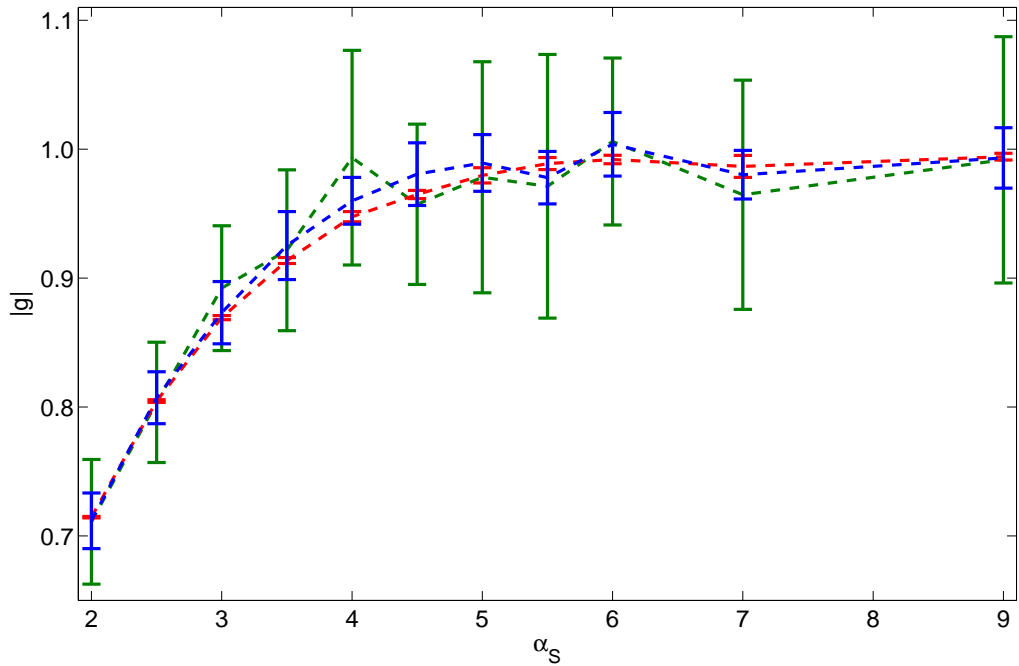}
\caption{Absolute value of steady-state gain from $\alpha_S$ to $\alpha_R$, estimated with the finite difference method. Shown are 95\% confidence intervals obtained with the method of batch means \cite{asmussen}. Green: CRN estimator. Blue: CFD estimator. Red: CRN combined with shadow functions.}
\label{SF_gain}
\end{figure}

The system reactions are given below:
\begin{equation}
\emptyset\xrightarrow{k_s}S\xrightarrow{k_d}\emptyset,~\emptyset\xrightarrow{k(S)}R\xrightarrow{1}\emptyset,
\end{equation}
where $k(S)=k_r/(S+K_m)$. The parameters used are $k_d=100$, $k_r=900$ and $K_m=0.9$, while $k_s$ is varied between 200 and 900 to study the effect of varying $\alpha_S:=\mathbb{E}_{\pi}[S]$ on $\alpha_R:=\mathbb{E}_{\pi}[R]$. More specifically (and similarly to Ref. \cite{warren12}), we want to calculate the gain
\[g=\frac{\partial\mbox{ln}(\alpha_R)}{\partial\mbox{ln}(\alpha_S)}=\frac{k_s}{\alpha_R}\frac{\partial\alpha_R}{\partial k_s}.\]
To this end we estimate $\partial\alpha_R/\partial k_s$ using finite differences with $\delta=2\cdot 10^{-2}k_s$ at several points between $k_s=200$ and $k_s=900$. Figure \ref{SF_gain} shows the calculated confidence intervals for $|g|$ obtained by the Common Random Number (CRN) estimator, the CRN estimator in conjunction with a shadow function and the CFD method. For each value of $k_s$, a simulated sample path of length $T=8000$ time units (t.u.) was used to generate 19 batches of length 400 t.u. each, while the first 400 t.u. were discarded as burn-in.

Shadow functions consisted of linear combinations of all monomials in two variables up to order three (that is, $\psi_i=S^j\cdot R^k$, with $0<j+k\leq 3$), together with \footnote{This is an example where the probabilistic interpretation of the Poisson equation given in subsection \ref{implementation} can provide useful intuition for the selection of approximating functions. In the case at hand, $f=R$, so $g$ (the solution to the Poisson equation) is expected to grow only very slowly with $S$, as the production rate of $R$ tends to zero as $S\to\infty$.} $\log(S+2)$. This set of $\psi_i$'s was selected manually and is definitely not the ``optimal'' choice. The training set used for regression consisted of all unique points visited by the process sample paths after a burn-in period. Two alternative weighting schemes were tested for each value of $k_s$: according to the first, all points were assigned equal weight ($M=I$), while in the second one the points were weighted according to the empirical distribution of the process, calculated using the simulated sample paths ($M\approx diag(\pi))$. Both schemes lead to variance reduction, and calculation of $\theta^*$ in each case can be performed very fast ($\sim 0.15$ sec), given the small number of training points ($\sim 2000$).

Post-processing of the trajectories for the evaluation of the shadow function over the different batches takes another 5 sec of CPU time. On the other hand, SSA simulation takes on average 40 sec, which demonstrates that the overhead associated with the shadow function usage is relatively small, while the computational savings in the estimation of $\alpha_R$ are significant, as Table \ref{VR_aR} demonstrates. Finally, a CFD simulation of the same length requires 220 sec of CPU time on average, while achieving a smaller magnitude of variance reduction.

\begin{table}[h!bt]
\renewcommand{\arraystretch}{1.3}
\caption{Variance reduction in the estimation of $g$}
\label{VR_aR}
\centering
\begin{tabular}{|c|c|c|c|c|c|c|c|c|c|c|c|}
\hline
$k_s/k_d$&2&2.5&3&3.5&4&4.5&5&5.5&6&7&9\\ \hline
CFD: $(\sigma_1^2/\sigma_2^2)$&5.0&5.3&4.0&5.6&21.0&6.5&16.7&25.3&6.9&22.2&16.6\\
CRN+SF: $(\sigma_1^2/\sigma_2^2)$&7464&2051&968&735&444&379&232&475&386&108&129\\
\hline
\end{tabular}
\end{table}

Before we leave this example, we should point out that application of the shadow function method to just the birth-and-death process of $S$ results in tremendous variance reduction for $f(S)=S^n$ and $n\leq 3$. As an example, Table \ref{poisson} shows the confidence intervals of uncentered moment estimates obtained with and without a shadow function for $k_s=500$, $k_d=100$, using a simulated trajectory of $T=5000$ t.u. and 10 batches. We attribute this phenomenon to the fact that the chosen set of functions $\psi_i$ can approximate the true solution to the Poisson equation very closely for $n\leq 3$. Of course, it is known that $S$ has a Poisson stationary distribution, which makes the use of a moment estimator pointless in this case. However, this interesting observation provides some heuristic justification for using polynomial approximating functions $\psi_i$.

\begin{table}[h!bt]
\renewcommand{\arraystretch}{1.3}
\caption{Variance reduction for a Poisson stationary distribution}
\label{poisson}
\centering
\begin{tabular}{|c|c|c|c|}
\hline
Moment & True value & $\alpha_1$ C.I. & $\alpha_2$ C.I.  \\ \hline
$\mathbb{E}_\pi[S]$ & 5 & $5.0012\pm0.008$ & $5\pm7.8\cdot 10^{-15}$\\
$\mathbb{E}_\pi[S^2]$ & 30 & $30.0241\pm0.0721$ & $30\pm1\cdot 10^{-12}$\\
$\mathbb{E}_\pi[S^3]$ & 205 & $205.42\pm1.219$ & $205\pm3.3\cdot 10^{-11}$\\
$\mathbb{E}_\pi[S^4]$ & 1555 & $1561.5\pm18.88$ & $1555.5\pm1.05$\\
\hline
\end{tabular}
\end{table}

\subsection{A six-dimensional system}
Our second example is a system consisting of two interacting genes, $A$ and $B$. The product of gene $A$ forms homodimers, which repress the expression of $A$, as well as heterodimers with the product of $B$, which repress the expression of $B$. The system species are listed in Table \ref{species}, while Table \ref{reactions} displays the reaction scheme and propensities of our model. Note that several parameters are assumed to be the same for the two genes for simplicity. For the same reason, the gene states are omitted from the model.

\begin{table}[h!bt]
\renewcommand{\arraystretch}{1.3}
\caption{Molecular species of the two-gene system}
\label{species}
\centering
\begin{tabular}{|c|c|}
\hline
Name & Symbol \\ \hline
Gene A mRNA & $m_A$\\
Protein A monomer & $p_A$\\
Protein A dimer & $p_{A_2}$\\
Gene B mRNA & $m_B$\\
Protein B momomer & $p_{B_2}$\\
A-B dimer & $p_{AB}$\\
\hline
\end{tabular}
\end{table}

\begin{table}[h!bt]
\renewcommand{\arraystretch}{1.3}
\caption{Reactions and propensities}
\label{reactions}
\centering
\begin{tabular}{|c|c|}
\hline
Reactions & Propensities \\ \hline
$\emptyset\xrightarrow{\lambda_1}m_A\xrightarrow{\lambda_2}\emptyset$ & $\lambda_1=k_{r}\displaystyle\frac{\phi^4}{\phi^4+p_{A_2}^4}$, $\lambda_2=k_{dr}m_A$\\
$\emptyset\xrightarrow{\lambda_3}p_A\xrightarrow{\lambda_4}\emptyset$ & $\lambda_3=k_{p}m_A$, $\lambda_4=k_{dp}p_A$\\
$p_A+p_A\rates{\lambda_5}{\lambda_6}p_{A_2}$ & $\lambda_5=k_{1}p_A(p_A-1)$, $\lambda_6=k_2p_{A_2}$\\
$\emptyset\xrightarrow{\lambda_7}m_B\xrightarrow{\lambda_8}\emptyset$ & $\lambda_1=k_{r}\displaystyle\frac{\phi^2}{\phi^2+p_{AB}^2}$, $\lambda_2=k_{dr}m_B$\\
$\emptyset\xrightarrow{\lambda_9}p_B\xrightarrow{\lambda_{10}}\emptyset$ & $\lambda_9=k_{p}m_A$, $\lambda_10=k_{dp}p_A$\\
$p_A+p_B\rates{\lambda_{11}}{\lambda_{12}}p_{AB}$ & $\lambda_{11}=k_3p_Ap_B$, $\lambda_{12}=k_4p_{AB}$\\
\hline
\end{tabular}
\end{table}

The system comprises six molecular species interacting through twelve reactions \footnote{Note that all examples presented in Ref. \cite{warren12} consist of two-species systems and no more than four reactions}. Our goal is to estimate the sensitivity of the steady-state mean of $p_{AB}$ (the second repressor dimer), denoted by $\alpha_{AB}$, to small variations of each of the system parameters. Once more, we compare the behavior of the CRN steady-state estimator with and without a shadow function to the performance of the CFD method. Shadow functions for this system consisted of linear combinations of all monomials of state pairs up to order two. Since the number of unique points visited by this six-dimensional process during simulation was (expectedly) too large to be handled with the least squares method, 10000 points sampled uniformly at random from this set were used in the regression step.

For the finite difference method we perturbed each parameter $p$ by $\delta=10^{-2}\cdot p$ and estimated the 95\% confidence intervals of each sensitivity estimate using batch means with 24 batches of length 4000 time units each (with an additional 4000 t.u. for burn-in). Prior to parameter perturbations, the estimate of $\alpha_{AB}$ was calculated for
\begin{eqnarray*}
p_0&=&\begin{bmatrix}k_r&\phi&k_{dr}&k_p&k_{dp}&k_1&k_2&k_3&k_4\end{bmatrix}\\
&=&\begin{bmatrix}1&60&0.1&1&0.5&0.02&0.08&0.02&0.1\end{bmatrix}.
\end{eqnarray*}
The estimates and their corresponding variances were: $\alpha_1=64.46$, $\sigma_1^2=2.68$, $\alpha_2=65.31$ and $\sigma_2^2=3.5\cdot 10^{-3}$. The results of the sensitivity analysis are summarized in Table \ref{sens}. CRN sensitivity estimates and their associated confidence intervals not accurate enough to provide useful information. On the contrary, using a shadow function results in great improvements, as now the relative magnitudes and signs of the various sensitivity coefficients can be meaningfully compared to each other.

\begin{table}[h!bt]
\renewcommand{\arraystretch}{1.3}
\caption{Sensitivity estimates and associated confidence intervals for the two-gene system. Note that a reduction of a confidence interval by a factor $r$ requires a variance reduction by a factor $r^2$, which can be achieved by running simulations $r^2$ times longer.}
\label{sens}
\centering
\begin{tabular}{|c|c|c|c|}
\hline
\centering Normalized sensitivity coefficient & CRN 95\% CI & CFD 95\% CI & CRN+SF 95\% CI  \\ \hline
$\alpha_2^{-1}\cdot\partial\alpha_{AB}/\partial{k_r}$ & $2.461\pm 1.339$ &$1.048\pm0.104$& $0.982\pm0.034$ \\
$\alpha_2^{-1}\cdot\partial\alpha_{AB}/\partial{\phi}$ & $0.025\pm 0.020$&$0.013\pm0.002$& $0.013\pm0.0005$\\
$\alpha_2^{-1}\cdot\partial\alpha_{AB}/\partial{k_{dr}}$ & $1.853\pm 11.222$ &$-10.246\pm 1.453$&$-9.777\pm 0.232$\\
$\alpha_2^{-1}\cdot\partial\alpha_{AB}/\partial{k_{p}}$ & $1.894\pm 1.236$ &$1.020\pm0.122$ &$0.967\pm0.025$\\
$\alpha_2^{-1}\cdot\partial\alpha_{AB}/\partial{k_{dp}}$ & $0.513\pm 2.049$ &$-1.876\pm0.286$ &$-1.888\pm0.058$\\
$\alpha_2^{-1}\cdot\partial\alpha_{AB}/\partial{k_1}$ & $137.769\pm 123.344$ &$18.437\pm8.181$ &$22.374\pm3.383$\\
$\alpha_2^{-1}\cdot\partial\alpha_{AB}/\partial{k_2}$ & $21.675\pm 16.660$ &$-3.053\pm1.417$ &$-2.785\pm0.333$\\
$\alpha_2^{-1}\cdot\partial\alpha_{AB}/\partial{k_3}$ & $87.001\pm 62.754$ &$27.001\pm4.43$ &$23.351\pm1.704$\\
$\alpha_2^{-1}\cdot\partial\alpha_{AB}/\partial{k_4}$ & $8.003\pm 16.060$ &$-4.818\pm0.756$ & $-4.597\pm0.328$\\
\hline
\end{tabular}
\end{table}

The variance reduction method remains quite efficient computationally in this case as well: SSA simulation of a $10^5$ t.u. trajectory takes about 17 sec of CPU time, while calculation of $\theta$ requires 1 sec and post-processing of the sample path another 3 sec. At the same time, a CFD simulation of the same length requires 95 sec of CPU time on average, while failing to achieve a comparable level of variance reduction.

\section{Discussion}
We demonstrated the applicability of the powerful shadow function method to the problem of steady-state simulation of stochastic chemical kinetics. Our results suggest that a significant increase in the efficiency of a steady-state estimator is possible by only a small increase in its computational cost. The method can be applied to the steady-state estimation of practically any function of the process, and can thus provide improved estimates of high order (cross-)moments, as well as estimates of stationary probabilities for subsets of the process state space, by using set indicators as cost functions. The magnitude of variance reduction achieved by the shadow function method allows also the efficient and precise computation of steady-state parameter sensitivities using the finite difference method.

The comparison of the efficiency of this approach for providing steady-state sensitivity estimates with the one presented in \cite{warren12} is the topic of our ongoing work. It would also be instructive to assess the relative strengths and weaknesses of the LSTD approximation algorithm for optimizing the shadow function \cite{meyn07} and test its scalability with system size and number of approximating functions (note that only one-dimensional examples are treated in \cite{meyn07}).

The proposed workflow for arriving at a useful shadow function can be improved at several points, by drawing from the large literature on function approximation techniques, in order to enlarge its range of applicability and its accuracy. However, even a crude approach such as the one presented above seems to be sufficient for systems of practical interest.

\appendix
\section{TAVC for shadow function estimator}\label{app_A}
From \eqref{cltvar} and \eqref{ghat}, $\sigma_2^2=-2\langle Qg_2,g_2\rangle$, where $g_2$ solves the Poisson equation $Qg_2=-f_c-Q(\psi\theta)$. This implies that $g_2=g_1-\psi\theta$, where $g_1$ is the solution of the Poisson equation $Qg_1=-f_c$. The variance of the shadow function estimator thus becomes
\begin{align*}\sigma_2^2&=-2\langle -f_c-Q(\psi\theta),g_1-\psi\theta\rangle\\
&= -2\left[\langle -f_c,g_1\rangle+\langle f_c,\psi\theta\rangle -\langle Q(\psi\theta),g_1\rangle +\langle Q(\psi\theta),\psi\theta\rangle\right]\\
&= \sigma_1^2-2\left[\langle f_c,\psi\theta\rangle -\langle Q(\psi\theta),g_1\rangle +\langle Q(\psi\theta),\psi\theta\rangle\right].
\end{align*}

\bibliographystyle{acm}

\begin{thebibliography}{10}

\bibitem{anderson12}
{\sc Anderson, D.~F.}
\newblock An efficient finite difference method for parameter sensitivities of
  continuous time markov chains.
\newblock {\em SIAM Journal on Numerical Analysis 50}, 5 (2012), 2237--2258.

\bibitem{asmussen2}
{\sc Asmussen, S.}
\newblock {\em Applied probability and queues}.
\newblock Springer, 2003.

\bibitem{asmussen}
{\sc Asmussen, S., and Glynn, P.}
\newblock {\em Stochastic simulation: Algorithms and analysis}, vol.~57.
\newblock Springer, 2007.

\bibitem{bhattacharya}
{\sc Bhattacharya, R.~N.}
\newblock On the functional central limit theorem and the law of the iterated
  logarithm for markov processes.
\newblock {\em Probability Theory and Related Fields 60}, 2 (1982), 185--201.

\bibitem{gillespie07}
{\sc Gillespie, D.~T.}
\newblock Stochastic simulation of chemical kinetics.
\newblock {\em Annu. Rev. Phys. Chem. 58\/} (2007), 35--55.

\bibitem{glynn90}
{\sc Glynn, P.~W.}
\newblock Likelihood ratio gradient estimation for stochastic systems.
\newblock {\em Communications of the ACM 33}, 10 (1990), 75--84.

\bibitem{glynn96}
{\sc Glynn, P.~W., and Meyn, S.~P.}
\newblock {A Lyapunov bound for solutions of the Poisson equation}.
\newblock {\em The Annals of Probability 24}, 2 (1996), 916--931.

\bibitem{henderson2}
{\sc Henderson, S.}
\newblock {\em Variance Reduction Via an Approximating {M}arkov Process}.
\newblock PhD thesis, Dept. of Operations Research, Stanford University, 1997.

\bibitem{henderson1}
{\sc Henderson, S., and Glynn, P.}
\newblock Approximating martingales for variance reduction in markov process
  simulation.
\newblock {\em Mathematics of Operations Research 27}, 2 (2002), 253--271.

\bibitem{hordijk}
{\sc Hordijk, A., Iglehart, D.~L., and Schassberger, R.}
\newblock Discrete time methods for simulating continuous time markov chains.
\newblock {\em Advances in Applied Probability\/} (1976), 772--788.

\bibitem{makowski02}
{\sc Makowski, A.~M., and Shwartz, A.}
\newblock The poisson equation for countable markov chains: probabilistic
  methods and interpretations.
\newblock In {\em Handbook of Markov decision processes}. Springer, 2002,
  pp.~269--303.

\bibitem{meyn07}
{\sc Meyn, S.}
\newblock {\em Control techniques for complex networks}.
\newblock Cambridge University Press, 2007.

\bibitem{munsky06}
{\sc Munsky, B., and Khammash, M.}
\newblock The finite state projection algorithm for the solution of the
  chemical master equation.
\newblock {\em The Journal of Chemical Physics 124\/} (2006), 044104.

\bibitem{norris98}
{\sc Norris, J.~R.}
\newblock {\em Markov chains}.
\newblock Cambridge University Press, 1998.

\bibitem{paulsson00}
{\sc Paulsson, J., Berg, O.~G., and Ehrenberg, M.}
\newblock Stochastic focusing: Fluctuation-enhanced sensitivity of
  intracellular regulation.
\newblock {\em Proceedings of the National Academy of Sciences 97}, 13 (2000),
  7148--7153.

\bibitem{plyasunov07}
{\sc Plyasunov, S., and Arkin, A.~P.}
\newblock Efficient stochastic sensitivity analysis of discrete event systems.
\newblock {\em Journal of Computational Physics 221}, 2 (2007), 724--738.

\bibitem{rathinam10}
{\sc Rathinam, M., Sheppard, P.~W., and Khammash, M.}
\newblock Efficient computation of parameter sensitivities of discrete
  stochastic chemical reaction networks.
\newblock {\em The Journal of chemical physics 132}, 3 (2010).

\bibitem{ruess11}
{\sc Ruess, J., Milias-Argeitis, A., Summers, S., and Lygeros, J.}
\newblock Moment estimation for chemically reacting systems by extended kalman
  filtering.
\newblock {\em The Journal of chemical physics 135}, 16 (2011), 165102--165102.

\bibitem{sheppard12}
{\sc Sheppard, P.~W., Rathinam, M., and Khammash, M.}
\newblock A pathwise derivative approach to the computation of parameter
  sensitivities in discrete stochastic chemical systems.
\newblock {\em The Journal of chemical physics 136\/} (2012), 034115.

\bibitem{singh11}
{\sc Singh, A., and Hespanha, J.~P.}
\newblock Approximate moment dynamics for chemically reacting systems.
\newblock {\em Automatic Control, IEEE Transactions on 56}, 2 (2011), 414--418.

\bibitem{tweedie98}
{\sc Tweedie, R.~L.}
\newblock Truncation approximations of invariant measures for {M}arkov chains.
\newblock {\em Journal of applied probability\/} (1998), 517--536.

\bibitem{warren12}
{\sc Warren, P.~B., and Allen, R.~J.}
\newblock Steady-state parameter sensitivity in stochastic modeling via
  trajectory reweighting.
\newblock {\em The Journal of Chemical Physics 136\/} (2012), 104106.

\end{thebibliography}

\end{document}